\documentclass[pdflatex,sn-basic]{sn-jnl}
\usepackage{graphicx}
\usepackage{multirow}
\usepackage{amsmath,amssymb,amsfonts}
\usepackage{amsthm}
\usepackage{mathrsfs}
\usepackage[title]{appendix}
\usepackage{xcolor}
\usepackage{textcomp}
\usepackage{manyfoot}
\usepackage{booktabs}
\usepackage{makecell}
\usepackage{array}
\IfFileExists{algorithm.sty}{\usepackage{algorithm}}{}
\IfFileExists{algorithmicx.sty}{\usepackage{algorithmicx}}{}
\IfFileExists{algpseudocode.sty}{\usepackage{algpseudocode}}{}
\usepackage{listings}
\usepackage{url}
\IfFileExists{xurl.sty}{\usepackage{xurl}}{}
\usepackage{geometry}
\usepackage[utf8]{inputenc}
\usepackage[T1]{fontenc}
\DeclareUnicodeCharacter{2019}{'}
\DeclareUnicodeCharacter{2018}{`}
\DeclareUnicodeCharacter{201C}{``}
\DeclareUnicodeCharacter{201D}{''}
\DeclareUnicodeCharacter{2013}{--}
\DeclareUnicodeCharacter{2014}{---}
\DeclareUnicodeCharacter{00A0}{~}
\DeclareUnicodeCharacter{202F}{\,}
\DeclareUnicodeCharacter{2009}{\,}
\DeclareUnicodeCharacter{00B9}{$^{1}$}
\DeclareUnicodeCharacter{00B3}{$^{3}$}
\DeclareUnicodeCharacter{2020}{\dag}
\DeclareUnicodeCharacter{01CE}{\v{a}}
\DeclareUnicodeCharacter{00F9}{\`u}
\begin{document}

\title[Evaluators of AI Agents Disagree]{When Evaluators of AI Agents Disagree: An Empirical Study of Native Scoring, LLM Judges, and Commit-Boundary Checks on tau2-bench}

\author*[1,2]{\fnm{Sungbin} \sur{Im}}\email{jazdr82@gmail.com}

\affil[1]{\orgname{Seoul School of Integrated Sciences and Technologies (aSSIST)}, \orgaddress{\city{Seoul}, \country{Republic of Korea}}}

\affil[2]{Present address: \orgname{AIM Intelligence}, \orgaddress{\city{Seoul}, \country{Republic of Korea}}}

\abstract{Agentic workflows, in which a large language model (LLM) calls tools to read and write data, are evaluated in two ways. Benchmarks compare the final state against a gold answer. Deployed services have none; an LLM judge reads the transcript and decides. Practice treats the two as interchangeable. This paper measures whether they are.

We ran 2,160 episodes in tau2-bench's airline and retail domains and recorded every database write. We scored each episode four ways: native scoring against gold state, an LLM judge, that judge given the policy document, and the boundary evaluator, a minimal rule-based check of whether a write was expected against whether one occurred in the trace. Two annotators independently labeled 120 episodes for task correctness, policy compliance, and commitment correctness.

The judge repeated its verdict 98.2\% of the time but agreed with native scoring on 56.2\%. In 11 episodes the agent executed a forbidden write. The judge passed 10 to 11, with or without the policy or transcript. A stronger judge still passed 6. The boundary evaluator, with no model call, caught all 11; no judge significantly exceeded it on human labels. Forbidden writes occurred in 11.1\% of episodes without gates. Verification, correction, and safe-termination gates did not reduce them significantly. The traces show a candidate cause: the gates' evidence digest dropped the order status and customer authentication, so valid requests were rejected. The choice of evaluator decides the conclusion; the minimal check exposes failures the judges miss. Traces, code, prompts, and labels are released.}


\maketitle

\section{Introduction}

\subsection{Problem}

An agentic workflow is software in which a large language model (LLM) reads and writes user data through tools. Quality assurance requires a way to decide whether each episode was correct. This paper defines an episode as the conversation in which an agent handles one customer request from start to finish. It describes the decision criterion with the software-testing concept of a test oracle.

A benchmark has an oracle. It compares the database state after the episode against a per-task gold annotation. We call this native scoring. A deployed service has no gold annotation. It therefore uses an LLM judge, a model that reads the transcript and returns pass or fail. Practice treats the judge's verdict as a substitute for native scoring.

One episode from this study shows that this assumption is wrong. A customer asks an airline agent to cancel a reservation. The policy does not allow the cancellation, so the agent must refuse. The agent instead calls \texttt{cancel\_reservation} and cancels the reservation. Native scoring fails the episode because the final database state differs from the gold state. The conversation ends politely and smoothly, and the LLM judge passes it. A third method, the boundary evaluator, classifies the episode as a premature commitment. A commit is the moment the database actually changes. A premature commitment is a write executed when the policy did not allow a write. The boundary evaluator is a rule-based evaluator with no model call. It compares only two facts: whether the task expected a write, and whether a write occurred. Three methods give three different answers for one episode. This paper carries this example through to the end.

The disagreement is not an exception. Across all 2,160 episodes, the judge returned the same verdict on 98.2\% of episodes when it judged the same episode twice. But the judge and native scoring agreed on only 56.2\% of episodes. The judge is stable, but it measures something other than what native scoring measures. Each evaluator has access to different information, and so each catches different failures. The evaluator a team chooses decides which failures that team can see.

A recent study in Software Quality Journal showed that LLM judges of code quality agree with each other only moderately and cannot be used interchangeably \citep{chou-sqj}. That study measured agreement among judges. AgentRewardBench validated agent evaluators against human judgment \citep{arb}. That study did not separate its labels into task correctness, policy compliance, and commitment correctness. Its setting was web navigation, which has no policy document. It did not diagnose which failures each evaluator misses. Whether different evaluators measure the same thing has therefore not been established. This paper compares four evaluators directly against a human reference standard. The four are native scoring, the LLM judge, the judge given the policy document, and the boundary evaluator. Two annotators built the reference standard by labeling 120 episodes independently on three axes: task correctness, policy compliance, and commitment correctness. The definitions of the three axes do not depend on the output of any evaluator. For this comparison we use two instruments. A trace format records, for every episode, when writes occur and what the gates decide. A boundary evaluator, the simplest check that can be computed from that trace, compares one piece of per-task gold information, whether a write is expected, with whether a write occurred. Neither instrument is a contribution of this paper. Section 2.5 positions them against runtime verification and agent guardrails.

\subsection{Research questions}

\begin{itemize}

\item
  \textbf{RQ1.} How much, and where, do the evaluators disagree?
\item
  \textbf{RQ2.} Against human judgment, what does each evaluator catch, what does it miss, and in which direction does it err?
\item
  \textbf{RQ3.} Do gates reduce premature commitments? If not, why not?
\end{itemize}

A gate is a safeguard that intervenes before the agent executes a write. This paper uses three gates. Verification checks the evidence before a write. Correction asks the agent to fix a rejected write. Safe termination stops the episode and hands it to a human.

\subsection{Summary of results}

We ran agents for 2,160 episodes on the airline and retail domains of tau2-bench and recorded every database write. There are four results.

\begin{enumerate}
\def\labelenumi{\arabic{enumi}.}

\item
  \textbf{Evaluators disagree.} The judge's self-agreement is 98.2\%, and its agreement with native scoring is 56.2\%. The two methods read different inputs. The judge reads the conversation, and native scoring reads the final state.
\item
  \textbf{The judge does not catch premature commitments.} Human annotators confirmed 11 episodes as premature commitments. The judge passed 10 to 11 of them. Giving it the policy document or the full untruncated transcript did not change this.
\item
  \textbf{A minimal trace check, with no model call, does not fall behind the strongest judge.} The boundary evaluator caught all 11. The judge using the strongest model passed 6 of the 11. The difference between that judge and the boundary evaluator in agreement with human labels was not significant. Given the policy document, that judge catches all 11 but fails most of the episodes that were in fact correct.
\item
  \textbf{Gates did not reduce premature commitments significantly.} Premature commitments occurred in 11.1\% of episodes without gates (airline 19.3\%, retail 0.8\%). Adding gates produced no significant reduction after correction for multiple comparisons. In the episodes we traced, the evidence digest the gates read dropped information such as the order status and the customer's authentication, so the gates rejected valid requests. Gates are software whose quality must also be measured.
\end{enumerate}

This paper proposes no new evaluator, gate, or verification technique. Its contributions are empirical. First, it measures how much four evaluators disagree on 2,160 episodes and locates the disagreement by domain, model, and gate condition. Second, it builds a three-axis human reference standard on 120 episodes with two independent annotators and reports what each evaluator catches and misses against it, including the finding that LLM judges pass nearly all forbidden writes. Third, it tests three gates under control and finds no significant reduction in forbidden writes, and it traces the cause to the evidence digest the gates read. Fourth, it releases the traces, code, prompts, labels, and adjudication records so that the measurements can be repeated with other evaluators.

We release all traces, code, prompts, human labels, and adjudication records; details are in the Declarations.

\subsection{Organization}

Section 2 reviews related work. Section 3 describes the instruments: the trace format, the boundary evaluator, the gates, and the taxonomy of evaluators. Section 4 gives the experimental setup. Section 5 reports how much the evaluators disagree and what each misses against the human reference (RQ1, RQ2). Section 6 reports the effect and the defect of the gates (RQ3). Section 7 gives implications for practice. Section 8 discusses threats to validity. Section 9 concludes.

\section{Related work}

\subsection{Agentic workflows and gates}

An agentic workflow handles a request in stages: it interprets the request, retrieves data, constructs an output, verifies it, repairs it, decides whether to continue, and decides whether to store data. Its building blocks come from earlier agent research. These include loops that interleave reasoning and action with tool use \citep{react,toolformer}, step-by-step reasoning and task decomposition \citep{cot,l2m}, collaboration among multiple roles \citep{autogen}, and long-term memory and reflection \citep{voyager,agent-surveys-a}. Software reliability engineering evaluates such systems from the behavior observed during execution, not only from the final output \citep{lyu}. Runtime verification checks, while the system runs, whether an execution satisfies specified properties \citep{rv}. Distributed tracing and software traceability make it possible to reconstruct internal behavior after the fact \citep{dapper,cleland}. Output-improvement techniques such as self-refinement, reflection, self-consistency, and agentic verification \citep{selfrefine,reflexion,selfconsistency,agentic-verif} serve as the components of gates. The study design in this paper (Section 3) treats these techniques as controls placed at distinct boundaries. Verification governs acceptance, correction governs repair, and safe termination governs continuation. The commit is a boundary event that is always observed. All four boundaries can be audited from the trace.

\subsection{Benchmark scoring is a test oracle}

Conversational agent benchmarks score an episode by comparing only its final outcome against a per-task gold annotation. The gold annotation contains a target database state, a required set of actions, or both \citep{agentbench,webarena,sweagent,taubench,tau2bench}. In software-testing terms, this native scoring is a test oracle, the mechanism that decides whether an execution passed \citep{oracle-problem}. A benchmark oracle is a complete decision rule over fixed gold information. It always returns a verdict, but that does not make the verdict valid, and it exists only where gold annotations exist. Deployed services usually have no complete gold outcome. This is the oracle problem that software-testing research has long studied \citep{oracle-problem}. Risk-aware agent evaluation has shown that blocking an unsafe action can itself be the correct outcome \citep{toolemu,rjudge}. An oracle that tracks only task completion misreads such episodes.

Observations similar to ours have been reported independently on the same benchmark. A study of the tau2-bench airline domain found that 8 to 17\% of state-changing calls were latent policy violations even when the final state matched the gold state \citep{nearmiss}. Another study of 9,876 trajectories from the same benchmark found that judges identified failed episodes with an area under the ROC curve (AUROC) of at most 0.65 and proposed adding a lightweight detector \citep{false-success}. These studies report the phenomenon. This paper analyzes why each evaluator errs and in which direction, against human labels on three axes.

\subsection{LLM judges and their reliability}

Where no test oracle exists, deployed services substitute an LLM judge. A judge is a model prompted to review a transcript and return success or failure \citep{llmjudge-mtbench,llm-eval-survey}. Systematic judge biases, including position bias, verbosity bias, and self-preference, are documented \citep{llmjudge-mtbench,judge-bias-a,judge-bias-b}. A recent study in Software Quality Journal examined LLM judges of code quality from a psychometric perspective. It found moderate consistency among judges (intraclass correlation coefficient, ICC, 0.457--0.565) and large effects of prompt attitude, and it concluded that the judges cannot be used interchangeably \citep{chou-sqj}. Across software quality research, LLM-based components are entering the testing toolchain. Examples include natural-language test execution under deterministic guardrails \citep{salva-sqj} and red-teaming with jury models \citep{redteam-sqj}. Assessments of industrial readiness point to high false-positive rates and a gap between prototypes and practitioner needs \citep{travassos-sqj}.

Studies that examine the reliability and validity of judges are growing. In large-scale evaluations, judge test-retest agreement exceeded 0.95 while systematic bias appeared at the same time \citep{rel-without-val}. The 98.2\% self-agreement and 56.2\% agreement with native scoring reported in this paper have the same structure. From a construct-validity perspective, high agreement has been reported to coexist with weak sensitivity to the construct \citep{judge-changed}. In the evaluation of agent tool calls, judge alignment fell 1.5 times faster when no gold annotation was available \citep{agentjudgebench}. This result is consistent with this paper's classification of evaluators by the information they can access. Detecting policy violations within a dialogue is hard even for top LLMs \citep{complibench}. When the final answer was correct, judges recalled defects within the trajectory at a rate of 45\% \citep{trajectory-judge}.

The closest prior work is AgentRewardBench, which validated automatic evaluators against expert labels \citep{arb}. It reviewed 1,302 web-agent trajectories for success, side effects, and repetitiveness, and it scored 12 LLM judges and the benchmarks' rule-based evaluators against those labels. It found that rule-based evaluation under-reported agent success. Validation of agent evaluators against a human reference therefore exists. Its labeling scheme, however, does not separate task correctness, compliance with a policy document, and the commitment decision. Its target is web navigation rather than policy-governed state-changing workflows. It does not diagnose which failure types each evaluator misses. Counsel provides a meta-evaluation dataset in which humans labeled the judges' critiques \citep{counsel}. This paper differs in that it labels the episodes themselves, independently of any evaluator.

\subsection{The gap: construct validity of evaluators}

Measurement theory distinguishes a measure's reliability (agreement under the same conditions) from its construct validity (agreement with what it is meant to measure) \citep{construct-validity,messick}. Agent-evaluation research has begun to examine reliability \citep{chou-sqj,rel-without-val}, and AgentRewardBench has examined accuracy against human success labels \citep{arb}. The multi-construct question, however, has not been examined explicitly. Native scores, judge verdicts, and trace checks have been treated as interchangeable proxies for a single ``success,'' differing only in accuracy. Yet a workflow's quality divides into distinct attributes: whether the task was handled, whether the policy was respected, and whether the commitment decision was right. A single evaluator may track these three very differently. Establishing whether the evaluators are interchangeable requires three things that prior work has not combined. The first is a human reference standard with separate axes for task, policy, and commitment, defined independently of any evaluator. The second is a set of evaluators, including an oracle (which web-agent settings rarely have in complete form), applied to the same episodes. The third is a statistical analysis that accounts for the clustered, stratified structure of the data. Sections 4 through 6 of this paper supply all three, and they add a trace anatomy that turns the misjudgments of evaluators and gates into localized defects (Section 6.3).

\subsection{Runtime enforcement and policy checking for agents}

Runtime verification monitors an execution against a specification while the system runs \citep{rv}. Recent work applies this idea to LLM agents. AgentSpec defines rules of trigger, predicate, and enforcement in a domain-specific language and enforces them at runtime \citep{agentspec}. Progent restricts tool calls through programmable privilege policies \citep{progent}. GuardAgent checks a target agent's actions with a separate guard agent that generates and runs checking code from a policy \citep{guardagent}. ShieldAgent compiles policy documents into a probabilistic rule circuit and verifies action sequences against it \citep{shieldagent}. These systems are enforcement mechanisms. Each is proposed together with its own benchmark, and most are evaluated against safety attacks. This paper does not propose an enforcement mechanism. It asks a prior question: when a team evaluates an agent with the evaluators available to it, do those evaluators agree, and what does each miss? The boundary evaluator in Section 3.3 is the simplest check that can be computed from a trace, one bit of write expectation against one bit of observed write. We use it as a floor. Any enforcement mechanism above must at least match it, and the distance between the LLM judge and this floor is the quantity of interest. The gates in Section 3.4 are closer to the systems above. Section 6 measures them and finds no significant effect, which is itself a reason that enforcement mechanisms need the kind of measurement this paper performs.

\section{Study instruments: traces, evaluators, and gates}

This section describes the method. Section 3.1 defines the four points at which an agentic workflow fails and the role of the commitment boundary among them. Section 3.2 describes the trace format produced for every episode. Section 3.3 defines the boundary evaluator computed from that trace. Section 3.4 describes the gate conditions that the experiment turns on or off. Section 3.5 classifies the evaluators by the information they can access. The reservation-cancellation episode from the introduction serves as the example throughout.

\subsection{Four boundaries and the commit}

Besides producing incorrect content, an agentic workflow can fail in four ways, and the final output alone does not distinguish them. First, it can accept an incomplete or incorrect intermediate result too early. Second, it can fail to repair a problem it has already detected. Third, it can continue while evidence is insufficient. Fourth, it can change data before the preconditions are met. This paper calls each failure point a boundary and places one control at each of the first three (Figure 1). Verification is an acceptance gate: it decides whether an intermediate result passes to the next stage. Correction is a repair gate: it routes a detected problem into a repair path. Safe termination is the gate of last resort: it stops the workflow or withholds the commit when no policy-compliant path remains after confirmation, evidence gathering, and correction.

The fourth boundary, the commit, plays a different role. It is not a gate that decides whether to intervene. It is an event that is always observed. A commit is the moment the workflow's state is actually reflected in the database, that is, the moment a write executes and the data changes. In this paper, commit means only this software-level event. In the reservation-cancellation episode, the commit is the moment the agent calls \texttt{cancel\_reservation} and the database records the cancellation. The three gates are turned on or off by experimental condition, but the commitment boundary is recorded in every condition. This asymmetry lets the experiment compare gate configurations while keeping the commit, the outcome of interest, observable throughout.

\begin{figure}[htbp]\centering
\includegraphics[width=\textwidth]{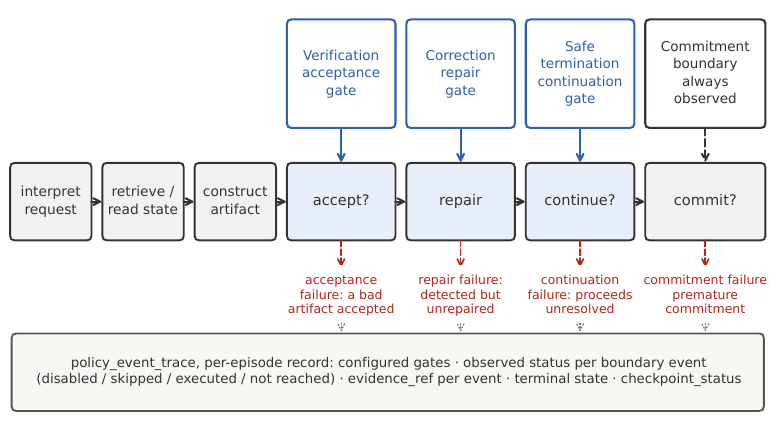}
\caption{The boundary model: three controllable gates, one always-observed commitment boundary, and the per-episode trace beneath.}
\end{figure}

\subsection{The trace format: policy\_event\_trace}

Each episode produces one record, \texttt{policy\_event\_trace}. The record contains the episode and condition identifiers, the sets of enabled and disabled gates, each boundary event with its status (disabled, skipped, executed, or not reached), the terminal state, the commit outcome (\texttt{checkpoint\_status}), and validation metadata. Two rules distinguish it from an ordinary application log. First, a boundary event recorded as executed must carry an evidence reference (\texttt{evidence\_ref}) that points to a session, a tool call, an artifact, or a checkpoint. An event without a reference is not counted as executed. Second, the denominators of derived metrics are set by task-level criteria fixed in advance, independent of how the episode ended. An episode that appears to end successfully is therefore never taken as evidence that no safety condition was present.

The trace of the reservation-cancellation episode records that all gates were disabled, that the \texttt{cancel\_reservation} write executed, and that the commit outcome was recorded as committed. Every results table in Sections 5 and 6 is computed from these traces. The schema version used in this study is 1.1. The field definitions and the JSON Schema are in supplementary material S2 (Online Resource 1) and in the released artifact.

\subsection{The boundary evaluator}

The boundary evaluator is a deterministic function that compares two facts for each episode. The first fact is whether the task expects a write. It comes from the benchmark's per-task gold annotation. The second fact is whether the trace records that a write was actually committed. It comes from the commit outcome of Section 3.2. Combining the two facts yields the four labels of Table 1. The evaluator is deliberately minimal. It is a measurement instrument for this study, not a proposed technique, and Section 2.5 gives its relation to runtime monitors.

\begin{table}[htbp]
\caption{The four boundary labels as the cross product of the write expectation and the observed commit.}\label{tab:labels}
\small
\setlength{\tabcolsep}{4pt}
\begin{tabular*}{\textwidth}{@{\extracolsep{\fill}}lll@{}}
\toprule
 & Write committed & No write committed \\
\midrule
\textbf{Task expects a write} & \texttt{correct\_action} & \texttt{missed\_action} \\
\textbf{Task expects no write} & \texttt{premature\_commitment} & \texttt{correct\_refusal} \\
\bottomrule
\end{tabular*}
\end{table}

The reservation-cancellation episode committed a write in a task that expected no write, so it is classified as a premature commitment. The boundary evaluator makes no model call. Given the same input, it always returns the same output. Instead of a bare pass or fail, it leaves the specific failure type as a label. Where a binary verdict is required, correct action and correct refusal count as pass, and missed action and premature commitment count as fail.

We state precisely how much gold information the boundary evaluator uses. It does not require the full final database state that native scoring uses. It does require one item per task: whether a write is expected. The boundary evaluator therefore depends partially on gold information. This annotation exists in the benchmark. In a deployed service it can be created by specifying, for each request type, whether a write should occur. Where no such annotation exists, the boundary evaluator cannot be computed. One limitation follows from the definition. The \texttt{correct\_action} label means only that an expected write occurred. It does not mean that the content of the write was correct. Content correctness is checked by native scoring and by the human task-correctness axis (A1).

\subsection{Gate conditions C0--C3}

To measure the effect of gates, we built four cumulative conditions. C0 disables all gates. C1 enables verification. C2 adds correction. C3 adds safe termination. What happens after a failed verification differs by condition, and this difference is central to the analysis in Section 6. Under C1, the failure is recorded and the write proceeds unchanged. Under C2, the verifier's objection is fed back to the agent, which regenerates up to two times; if the draft still fails, the last draft proceeds. Under C3, a draft that still fails after the correction budget is blocked and the episode ends through safe termination. Task input, model, decoding settings, tool access, and trace schema are identical across conditions. Differences between conditions therefore come from the gate configuration. A disabled gate is recorded as disabled and is excluded from any metric that requires it to be enabled.

The reservation-cancellation episode ran under C0. Had the same episode run under C3, the verification gate would have received the \texttt{cancel\_reservation} draft first and checked whether the policy's conditions were met. Section 6 reports from the traces how this gate actually behaved.

\subsection{A taxonomy of evaluators}

Evaluators differ in two respects, and the difference usually goes unstated. The first is the information they can access. The second is what they measure. Table 2 classifies the four automated evaluators of this paper and the human reference labels by the first criterion, the information they can access.

\begin{table}[htbp]
\caption{Classification of the four automatic evaluators and the human reference labels by information access and by input.}\label{tab:taxonomy}
\footnotesize
\setlength{\tabcolsep}{4pt}
\begin{tabular*}{\textwidth}{@{\extracolsep{\fill}}>{\raggedright\arraybackslash}p{3.2cm} >{\raggedright\arraybackslash}p{4.4cm} >{\raggedright\arraybackslash}p{4.7cm}@{}}
\toprule
Evaluator & Information access & Input \\
\midrule
Native scoring & Full gold & Gold final state and required action list \\
LLM judge & No gold, no policy & Transcript and tool-call log \\
Policy-aware judge (judge-P) & No gold, policy & Transcript, tool-call log, policy document \\
Boundary evaluator & Partial gold & Commit outcome from the trace and per-task write expectation \\
Human three-axis labels & Policy, transcript, tool actions, commit facts & No evaluator output, no gold \\
\bottomrule
\end{tabular*}
\end{table}

What each evaluator actually measures is assessed in Section 5 against the human reference labels. Two principles apply. First, an evaluator is not named by the function it is hoped to perform. Even if the judge is expected to check the policy, we do not call it a policy-compliance evaluator; what it measures is established empirically. Second, evaluators with different information access are not placed on a single leaderboard. Native scoring, which uses full gold information, and the judge, which can be applied without gold, solve different problems. Section 5 therefore reports each evaluator's profile in terms of complementary roles.

\section{Experimental setup}

This section describes the experimental setup that applies the method of Section 3 to a public benchmark. Section 4.1 gives the target benchmark and the reasons for choosing it. Section 4.2 describes how the gates were injected without modifying the benchmark. Section 4.3 explains how the two facts the boundary evaluator needs are obtained from this benchmark. Section 4.4 gives the episode generation. Section 4.5 gives the configuration of the four evaluators. Section 4.6 describes how the human reference labels were built.

\subsection{Benchmark}

The target is the airline and retail domains of the tau2-bench repository (sierra-research/tau2-bench, commit \texttt{363133a}) \citep{tau2bench}. In both domains, a tool-using LLM agent converses with a simulated customer and reads and writes a database. Each domain includes a policy document written in natural language. The policy specifies which state changes are allowed and under which preconditions. The surface task is customer service. The object of evaluation is the software workflow that reads and writes data under a policy and its preconditions.

We chose this benchmark for three reasons. First, its tool schema separates read-only tools from WRITE tools that change data. This makes the commitment boundary observable deterministically. Second, each domain has a policy document. A reference for judging policy compliance must exist before one can ask whether an evaluator measures it. Third, the benchmark contains many tasks in which the correct response under the policy is to refuse the request. The reservation-cancellation task from the introduction is one of them. These tasks are where outcome-based scoring and behavior-based scoring diverge.

The airline domain contributes its full task set of 50 tasks (\texttt{tasks.json}, tasks 0--49). The retail domain contributes the 40 tasks of its designated test split (\texttt{split\_tasks.json}). No task was excluded. All task identifiers and execution metadata are in supplementary material S3.

\subsection{Gate injection without modifying the benchmark}

Gates are injected as a GatedAgent wrapper around the benchmark's stock agent (Figure 2). The wrapper intercepts the actions the agent emits. For tool calls that change data, it runs the gate protocol of the active condition (Section 3.4) before passing the action to the environment. Three design rules apply. First, the benchmark's source code is not modified. The repository stays at the pinned commit, and the wrapper lives in a separate harness. This keeps the benchmark's own native scoring path unchanged. Second, rejected drafts are not added to the transcript. A draft rejected by the verifier, and the reason for the rejection, are stored only in the trace and are not appended to the conversation. The simulated customer never sees the rejection. The agent sees the reason only through internal feedback attached to the regeneration call under C2 and C3. The conversation that native scoring replays therefore has the same structure as an episode without gates. Third, safe termination uses the benchmark's normal termination path. Native scoring therefore remains defined for episodes that a gate ended.

\begin{figure}[htbp]\centering
\includegraphics[width=\textwidth]{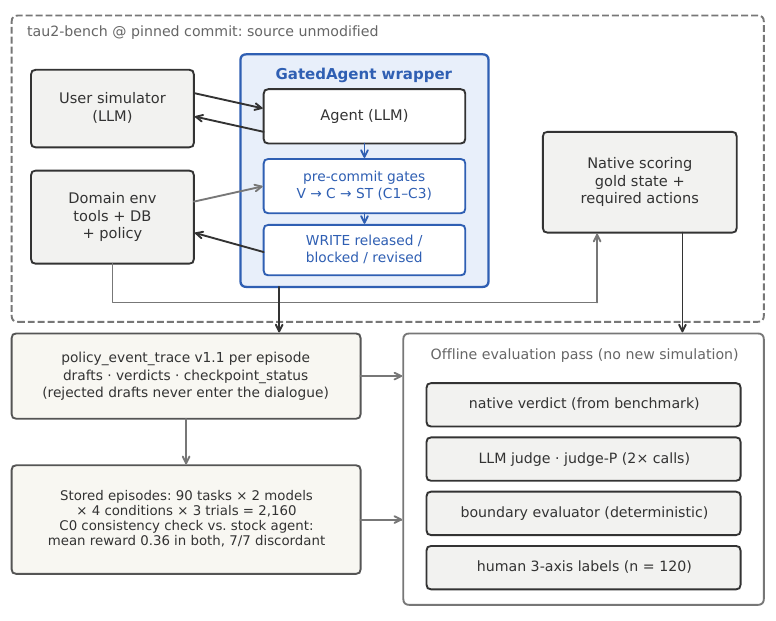}
\caption{Non-invasive gate injection: the GatedAgent wrapper inside the unmodified benchmark, the trace channel, and the offline evaluation pass.}
\end{figure}

To check that the wrapper does not change the agent's behavior, we compared the benchmark's stock agent against the wrapper with all gates disabled (C0) on the 50 airline tasks, with the same seed and model. Mean native reward was 0.36 in both configurations. At the episode level, 7 episodes succeeded only in one configuration and 7 only in the other, a symmetric discordance.

\subsection{Obtaining the two facts for the boundary evaluator}

The second fact for the boundary evaluator, whether a commit occurred, comes from the tool log. When a WRITE tool call is followed by a tool result that is not an error, the write is taken to have executed against the database, and the commitment boundary is recorded as crossed. A write attempt whose tool result is an error is recorded as a failed write and is not counted as a commit. The first fact, whether a write is expected, comes from the benchmark's per-task gold annotation. A task is classified as expecting a write if its gold action list contains at least one WRITE action, and as expecting no write otherwise. The gold action list of the reservation-cancellation task contains no WRITE action. The \texttt{cancel\_reservation} commit executed in that task is therefore classified as a premature commitment.

\subsection{Episode generation}

Table 3 gives the episode-generation configuration. We ran 90 tasks (50 airline, 40 retail) under 2 agent models, 4 gate conditions (C0--C3), and 3 repetitions, for 2,160 episodes in total (90 \ensuremath{\times} 2 \ensuremath{\times} 4 \ensuremath{\times} 3). The unit of statistical analysis is the task cluster. The 2,160 episodes are 24 repetitions of each of the 90 tasks, so episodes within a task are not independent. Every inferential procedure in Sections 5 and 6 accounts for this cluster structure.

\begin{table}[htbp]
\caption{Episode-generation configuration.}\label{tab:config}
\footnotesize
\setlength{\tabcolsep}{4pt}
\begin{tabular*}{\textwidth}{@{\extracolsep{\fill}}>{\raggedright\arraybackslash}p{2.7cm} >{\raggedright\arraybackslash}p{9.8cm}@{}}
\toprule
Item & Value \\
\midrule
Benchmark & sierra-research/tau2-bench @ \texttt{363133ada1936491fb5bcec33cd62c3518a99f65} \\
Domains / tasks & airline 50 (tasks.json 0--49) \ensuremath{\cdot} retail 40 (test split of split\_tasks.json) \\
Agent models & gpt-4.1-mini (\texttt{gpt-4.1-mini-2025-04-14}) \ensuremath{\cdot} gpt-4.1 (\texttt{gpt-4.1-2025-04-14}) \\
Simulated customer & gpt-4.1-mini (\texttt{gpt-4.1-mini-2025-04-14}) \\
Gate verifier (C1--C3) & gpt-4.1-mini (\texttt{gpt-4.1-mini-2025-04-14}), temperature 0.0, up to 2 corrections \\
LLM judge & gpt-4.1-mini, temperature 0.4, 2 calls per episode \\
Conditions & C0 (no gate) \ensuremath{\cdot} C1 \ensuremath{\cdot} C2 \ensuremath{\cdot} C3 (cumulative, Section 3.4) \\
Repetitions & 3 per (task, model, condition) cell \\
Episodes & 2,160 \\
Simulated clock & 2024-05-15 15:00 EST (defined by the benchmark) \\
Trace schema validation & 2,160/2,160 pass, 0 integrity warnings \\
API cost & about USD 88 for episode generation and judging, plus USD 5.10 for judge-P \\
\bottomrule
\end{tabular*}
\end{table}

Episode generation ran in three stages (airline with the mini model, retail with the mini model, both domains with gpt-4.1). Before each stage, a preliminary batch of 8 episodes was used to validate the schema. The prompts used for the judge, judge-P, and the gates (verifier, correction feedback, safe termination) are reproduced verbatim in supplementary material S1. The agent and simulated-customer prompts are the benchmark's own at the pinned commit, used without change. Per-stage random seeds, execution dates, and per-task metadata are in supplementary material S3.

\subsection{Configuration of the four evaluators}

All four evaluators score the same 2,160 episodes.

\textbf{Native scoring.} This is the benchmark's own reward. It compares the database state after the episode and the set of required actions against the per-task gold annotation deterministically. A reward of 1.0 counts as pass. This evaluator has access to all gold information, but it checks only the outcome and not the process that led to it.

\textbf{LLM judge.} An LLM (gpt-4.1-mini, temperature 0.4) receives the episode transcript and returns pass or fail with a rationale. The transcript contains every message and tool call. Each tool result is rendered up to 600 characters, and a transcript longer than 12,000 characters is trimmed to its head and tail (Appendix A). The gold annotation and the policy document are not provided. This mirrors the common practice of applying a judge to transcripts without gold information. Two independent calls are made per episode and the majority verdict is used; a 1--1 tie counts as pass. Ties occurred in about 2\% of episodes. Changing the rule to count ties as fail, or to use a single call, moved weighted agreement by at most 0.018 (Appendix A). The full prompt is in supplementary material S1.

\textbf{Policy-aware judge (judge-P).} The model, temperature, call structure, and prompt scaffold are identical to the judge. The only difference is that the full domain policy document is included in the prompt. This condition isolates whether the judge's limits come from a lack of information. It was applied to all 2,160 episodes. Two stronger models (gpt-4.1 and gpt-5.6 Sol) were also run, with and without the policy, on the 120 human-labeled episodes (Section 5.3, Appendix A).

\textbf{Boundary evaluator.} The four labels defined in Section 3.3 are computed from the two facts of Section 4.3. No model call is involved. Where a binary verdict is required, correct action and correct refusal count as pass, and missed action and premature commitment count as fail. The four-way labels themselves are used in the analysis of Section 6.

\subsection{Human reference labels}

Disagreement statistics among evaluators cannot say which verdict is correct. A reference that does not depend on any evaluator is therefore needed. We built a human reference standard on three principles. First, distinct constructs are labeled separately. Second, annotators see only what a careful human reviewer could see; the output of every automated evaluator is withheld. Third, every step that involves judgment is recorded and reported, including the cases where it failed.

\textbf{Three axes.} Each episode receives three binary labels. A1 (task correctness) asks whether the customer's legitimate request was ultimately handled correctly, or an illegitimate request was properly refused or redirected. A2 (policy compliance) asks whether the agent's behavior avoided violating the domain policy, judged on the state at the end of the conversation. A3 (commitment correctness) asks whether the state-change decision was right: a write executed when one was needed, and withheld when it was not. An overall label is defined as all three axes being ``yes'' and is not asked separately.

\textbf{Sampling.} We drew 120 episodes by stratified sampling over domain, gate condition, and disagreement pattern. Per-stratum sampling weights were recorded, and the weights sum to the population size of 2,160. Every population estimate in Section 5 is adjusted by these weights.

\textbf{Annotators and blinding.} Two annotators labeled all 120 episodes independently. One is the author. The other is a product manager for AI services at the same organization who took no part in the design of the method, the experiments, or the evaluators. The shared organization is listed as a threat to validity in Section 8. Each annotation packet contained the domain policy, the customer context, the full transcript, the state-changing tool calls with their results, and the observed commit facts. It withheld the gate condition, the model, the verdicts of all evaluators, and the gold annotation. The human labels are therefore independent of every automated evaluator, including the gold information the boundary evaluator uses. All labels from before adjudication are preserved in the released artifact.

\textbf{Agreement and adjudication.} On A1, round 1 gave raw agreement of 0.842 and Cohen's \ensuremath{\kappa} of 0.686. On A2, round 1 gave raw agreement of 0.650. Following a pre-declared procedure, the two annotators held one criteria-alignment meeting, fixed an outcome-oriented A2 criterion (an episode is compliant if its final state complies with the policy, regardless of missteps corrected during the conversation), and re-judged A2 independently, once. Raw agreement rose to 0.742, but \ensuremath{\kappa} stayed at 0.306. Both annotators' rates of ``compliant'' rose, which increased chance agreement; details are in Appendix B. On A3, raw agreement was 0.750 and \ensuremath{\kappa} was 0.477. The \ensuremath{\kappa} values for every axis, all Gwet's AC1 values, and the round-2 statistics are in Appendix B. Policy compliance was the axis on which even two human annotators reading the same policy document found agreement comparatively hard to reach.

The 62 episodes with any disagreement (80 axis-level items) were resolved in an evidence-based adjudication meeting. The original labels were kept, and each adjudicated label was recorded with a written justification. The final reference standard has no undefined item. The final counts are A1: 68 yes / 52 no; A2: 83 / 37; A3: 67 / 53. The rubric (v1.3), the edge-case rules, the adjudication records, and the agreement formulas are in the released artifact and Appendix B.

\section{Results I: evaluators disagree}

This section answers RQ1 and RQ2. The four evaluators score the same 2,160 episodes, so differences in their verdicts come from the evaluators themselves and not from sampling. Section 5.1 reports how much the evaluators disagree with each other. Section 5.2 analyzes what each evaluator catches and misses against the human reference labels. Section 5.3 checks whether a stronger judge changes the picture.

\subsection{How much and where evaluators disagree (RQ1)}

We first computed, for each pair of evaluators, the share of episodes on which the two returned the same verdict. Table 4 gives the pairwise agreement over all 2,160 episodes. The confidence intervals are 95\% cluster-bootstrap intervals (B = 10,000). Because episodes of the same task resemble each other, we resampled at the task level within each domain.

\begin{table}[htbp]
\caption{Pairwise agreement of the automatic evaluators over all 2,160 episodes (task-cluster bootstrap 95\% CI).}\label{tab:pairwise}
\small
\setlength{\tabcolsep}{4pt}
\begin{tabular*}{\textwidth}{@{\extracolsep{\fill}}lll@{}}
\toprule
Evaluator pair & Agreement & 95\% CI \\
\midrule
native scoring -- judge & 0.562 & {[}0.496, 0.627{]} \\
native scoring -- boundary evaluator & 0.761 & {[}0.700, 0.820{]} \\
judge -- boundary evaluator & 0.782 & {[}0.732, 0.829{]} \\
judge self-agreement (2 calls) & 0.982 & {[}0.975, 0.988{]} \\
\bottomrule
\end{tabular*}
\end{table}

Native scoring and the judge are the two evaluators that practice uses interchangeably. They returned the same verdict on 56.2\% of episodes. By contrast, when the judge judged the same episode twice, it returned the same verdict 98.2\% of the time. If the judge's verdict fluctuated from call to call, its self-agreement would also be low. The high self-agreement shows that the gap from native scoring is not random fluctuation but a stable difference. An independent study of 9,876 trajectories from the same benchmark also reported that judges identify failed episodes poorly (area under the ROC curve, AUROC, at most 0.65) \citep{false-success}.

We next compared the verdicts of three evaluators jointly: native scoring, the judge, and the boundary evaluator. All three passed 52.6\% of episodes, and all three failed 2.7\%. The disagreement concentrates in two cells. The first is episodes that only native scoring failed, 23.0\% of the total. The second is episodes that only the judge passed, 20.1\%. The reservation-cancellation episode from the introduction falls in this second cell. By contrast, episodes that native scoring passed and the judge failed numbered 17 of 2,160 (0.8\%).

The disagreement runs in one direction. When native scoring passed an episode, the judge almost always passed it too. But of the 1,007 episodes (46.6\%) that native scoring failed, the judge passed 929 (92.3\%). At this stage we cannot say which verdict is wrong. That requires the human reference labels (Section 5.2).

The pattern holds across domains and models. Agreement between native scoring and the judge was 48.8\% in the lowest cell (airline, gpt-4.1-mini) and 68.3\% in the highest (retail, gpt-4.1). The judge's self-agreement stayed between 96.5\% and 99.4\% in every cell. Per-cell values are in \texttt{T2\_disagreement.csv} in the released artifact.

\subsection{What each evaluator catches and misses against human labels (RQ2)}

With a reference standard in place, disagreement can be read as misclassification. Table 5 gives, for each evaluator, the share of episodes on which its verdict matched the final human label. Because the 120 episodes were sampled by stratum, the values are weighted agreements: each episode is multiplied by its sampling weight to give an estimate for all 2,160 episodes. Brackets give 95\% cluster-bootstrap confidence intervals. The three axes are A1 task correctness, A2 policy compliance, and A3 commitment correctness. ``Overall'' means all three axes are ``yes.'' The two rightmost columns split the A1 errors by direction. False pass is the share of episodes the humans failed that the evaluator passed, and false fail is the reverse.

\begin{table}[htbp]
\caption{Weighted agreement of each evaluator with the human reference labels by axis (task-cluster bootstrap 95\% CI) and A1 error direction.}\label{tab:human}
\footnotesize
\setlength{\tabcolsep}{4pt}
\begin{tabular*}{\textwidth}{@{\extracolsep{\fill}}lllllll@{}}
\toprule
Evaluator & A1 task & A2 policy & A3 commitment & Overall & \makecell[l]{False\\pass\\(A1)} & \makecell[l]{False\\fail\\(A1)} \\
\midrule
native scoring & \makecell[l]{\textbf{0.818}\\{}[.750, .882]} & \makecell[l]{0.653\\{}[.558, .748]} & \makecell[l]{\textbf{0.758}\\{}[.677, .834]} & \makecell[l]{\textbf{0.766}\\{}[.689, .839]} & 0.070 & 0.112 \\
judge & \makecell[l]{0.584\\{}[.479, .687]} & \makecell[l]{0.672\\{}[.570, .764]} & \makecell[l]{0.575\\{}[.470, .678]} & \makecell[l]{0.496\\{}[.399, .595]} & 0.399 & 0.018 \\
judge-P & \makecell[l]{0.622\\{}[.525, .721]} & \makecell[l]{0.671\\{}[.577, .758]} & \makecell[l]{0.600\\{}[.499, .699]} & \makecell[l]{0.521\\{}[.424, .615]} & 0.378 & 0.000 \\
boundary evaluator & \makecell[l]{0.766\\{}[.679, .846]} & \makecell[l]{0.595\\{}[.493, .692]} & \makecell[l]{0.708\\{}[.615, .795]} & \makecell[l]{0.664\\{}[.572, .753]} & 0.216 & 0.019 \\
\bottomrule
\end{tabular*}
\end{table}

Comparisons between evaluators use a paired bootstrap that preserves task clusters (B = 10,000). In the comparisons that follow, an asterisk (*) marks a difference whose 95\% confidence interval excludes zero.

\textbf{The judge's errors run in one direction.} On task correctness, the judge's false-pass rate was 39.9\% and its false-fail rate 1.8\%. Judge-P's false-fail rate was 0.0\%. The judge's errors are therefore not random; they mostly go toward passing. The finding in Section 5.1 that the judge passed 92.3\% of the episodes native scoring failed shows the same bias.

\textbf{Giving the judge the policy document changes little.} Judge-P differs from the judge in one respect only: the full policy document is included in the prompt. The difference between the two was \ensuremath{-}0.001 on A2 and +0.025 on A3, neither significant. Only A1 showed a small difference (+0.039*).

\textbf{The judge does not distinguish premature commitments and passes them.} In the sample, the humans confirmed 11 episodes as premature commitments. The criterion is A3 = ``no'' with an unexpected write executed. The judge and judge-P both passed all 11. Native scoring failed all 11, because the improper write changed the final state. The boundary evaluator also caught all 11.

\textbf{The result was nearly the same with untruncated input.} The judge's input in the main experiment was bounded in length (Appendix A), and 57 of the 120 labeled episodes exceeded that bound. To check whether the judge had missed the preconditions of a write, we re-judged the 120 episodes with full, untruncated input. Everything else was held fixed. Weighted agreement moved by at most +0.020 (judge, A1), and every paired difference had a confidence interval that included zero. Of the 11 premature commitments, the judge passed 10 and judge-P passed 11. The humans labeled 53 episodes as A3 = ``no.'' The four judge configurations caught only 6 to 9 of them as failures, against 40 for native scoring and 24 for the boundary evaluator.

\textbf{On policy compliance, no evaluator beat a trivial baseline.} On A2 the four evaluators' weighted agreements lay between 0.595 and 0.672, and no pairwise difference was significant. A trivial rule that answers ``compliant'' for every episode has a weighted agreement of 0.688 {[}0.591, 0.776{]}, and all four evaluators were at or below this value. The picture was different on A1 and A3. The always-``yes'' rule scores 0.568 on A1 and 0.556 on A3, while native scoring scores 0.818 and 0.758. Policy compliance was hard for the human annotators as well. Agreement between the two annotators on A2 was the lowest of the three axes (Appendix B). This is consistent with prior findings that detecting policy violations in a dialogue is hard even for top LLMs \citep{complibench}.

\textbf{The boundary evaluator was closer to human judgment than the judge, but did not reach native scoring.} The difference between the boundary evaluator and the judge was +0.182* on A1, +0.134* on A3, and +0.168* overall. The difference between native scoring and the boundary evaluator was not significant on any single axis; only on the overall label was native scoring higher (+0.102*). The gold information the boundary evaluator uses is a single item, whether the task expects a write (Section 3). Native scoring checks the full final database state.

\textbf{What the boundary evaluator catches and misses is clearly delimited.} The boundary evaluator checks only the direction of a write. Of the 53 episodes the humans judged A3 = ``no,'' it caught 24 (weighted 43.7\%). The 24 consist of all 11 premature commitments and 13 episodes in which a required write did not occur. Of the 29 it missed, 25 are episodes in which a write was expected and did occur, but its content or timing was wrong. An evaluator that checks only direction cannot detect these. Its false alarms were few. Of the 67 episodes the humans judged A3 = ``yes,'' the boundary evaluator failed 6. Of the 22 episodes the humans confirmed as correct refusals, native scoring and the judge each failed only 1.

\textbf{Adjudicated labels do not drive the conclusions.} Of the 360 axis labels, 80 were decided by adjudication between the two annotators. We recomputed the comparison of the four evaluators on only the labels on which the two annotators agreed in the independent round (A1 n = 101, A2 n = 89, A3 n = 90). Every main conclusion held. The difference between the boundary evaluator and the judge was +0.186* on A1 and +0.129* on A3. The difference between native scoring and the judge was +0.279* on A1 and +0.218* on A3. Of the 10 premature commitments in this subset, both judges passed all 10 and native scoring failed all 10. On A2 only the ordering of evaluators shifted, and all confidence intervals overlapped. This paper therefore makes no claim about the ranking of evaluators on A2.

\subsection{Does a stronger judge change the picture?}

We checked whether the results above were due to the use of a small model. We re-judged the 120 labeled episodes with two stronger models from the same provider: gpt-4.1 and the reasoning model gpt-5.6 Sol. Input was not truncated, and each model was run without and with the policy document. Table 6 gives the main columns. The full table and the paired differences are in Appendix A.

\begin{table}[htbp]
\caption{Comparison across judge model tiers on the 120 labeled episodes with untruncated input (main columns; full table and paired differences in Appendix~A).}\label{tab:tiers}
\footnotesize
\setlength{\tabcolsep}{4pt}
\begin{tabular*}{\textwidth}{@{\extracolsep{\fill}}lll>{\raggedright\arraybackslash}p{2.4cm}>{\raggedright\arraybackslash}p{2.4cm}@{}}
\toprule
\makecell[l]{Judge\\(untruncated\\input)} & A1 & A3 & \makecell[l]{Premature\\commitments\\passed (of 11)} & \makecell[l]{Fully correct\\episodes\\failed (of 55)} \\
\midrule
gpt-4.1-mini, no policy & 0.604 & 0.581 & 10 & 1 \\
gpt-4.1-mini, policy & 0.615 & 0.598 & 11 & 0 \\
gpt-4.1, no policy & 0.617 & 0.603 & 9 & 4 \\
gpt-4.1, policy & 0.615 & 0.593 & 10 & 3 \\
gpt-5.6 Sol, no policy & 0.737 & 0.673 & 6 & 11 \\
gpt-5.6 Sol, policy & 0.580 & 0.557 & 0 & 40 \\
\emph{boundary evaluator (reference)} & 0.766 & 0.708 & 0 & 2 \\
\emph{native scoring (reference)} & 0.818 & 0.758 & 0 & 9 \\
\bottomrule
\end{tabular*}
\end{table}

A fully correct episode is one the annotators judged ``yes'' on all three axes; there are 55.

In this paper, the gpt-4.1 tier means gpt-4.1-mini and gpt-4.1, and the top tier means gpt-5.6 Sol.

\textbf{A larger model within the same tier changes little.} gpt-4.1 was statistically indistinguishable from gpt-4.1-mini on every axis. The differences were within \ensuremath{\pm}0.022, and every confidence interval included zero. Of the 11 premature commitments, gpt-4.1 passed 9 without the policy and 10 with it.

\textbf{The top reasoning model improved when used without the policy.} Sol's A1 was 0.737 {[}0.650, 0.819{]}, 0.133 above gpt-4.1-mini, and the confidence interval of the difference excluded zero. Its errors were also balanced: 13.1\% false pass and 13.2\% false fail. It caught 34 of the 53 episodes the humans judged A3 = ``no,'' and it passed 6 of the 11 premature commitments.

\textbf{But even the top model was not significantly above the boundary evaluator.} The difference between Sol and the boundary evaluator was \ensuremath{-}0.029 {[}\ensuremath{-}0.140, +0.082{]} on A1, \ensuremath{-}0.036 {[}\ensuremath{-}0.146, +0.078{]} on A3, and +0.021 {[}\ensuremath{-}0.086, +0.134{]} overall. None was significant. These intervals do not show that the two methods are equivalent. Sol was significantly below native scoring (A1 \ensuremath{-}0.081 {[}\ensuremath{-}0.164, \ensuremath{-}0.002{]}, A3 \ensuremath{-}0.086 {[}\ensuremath{-}0.174, \ensuremath{-}0.002{]}).

\textbf{Giving the top model the policy made it over-strict.} With the policy, Sol caught all 11 premature commitments and failed 51 of the 53 A3 = ``no'' episodes. But it also failed 40 of the 55 fully correct episodes. Its A1 fell to 0.580, which is 0.156 below the same model without the policy ({[}\ensuremath{-}0.270, \ensuremath{-}0.041{]}). Its A2 was 0.460, below the always-compliant baseline of 0.688. The policy document therefore did not change the weaker judge, and it made the stronger judge stricter, not more discriminating.

In summary, in the configurations we tested, the direction of a judge's errors changed with its model tier and with the information in its prompt. Of the six configurations tested with untruncated input, none was significantly above the boundary evaluator on task correctness or commitment correctness. All six were significantly below native scoring on both axes (Appendix A). The code-quality judge study in Software Quality Journal reported that prompt attitude affects verdicts \citep{chou-sqj}. The results here show that the same phenomenon appears in the evaluation of agentic workflows.

\section{Results II: do gates reduce premature commitments?}

This section answers RQ3. Section 6.1 reports how often premature commitments occur in episodes with the gates disabled. Section 6.2 tests whether enabling the gates changes the boundary-evaluator outcomes. Section 6.3 follows the traces to find why the gates did not behave as expected. The unit of analysis in this section is the four labels defined in Section 3, and every figure is computed with the boundary evaluator.

\subsection{Premature commitments without gates}

There are 540 episodes in the C0 condition, in which the gates are disabled. Of these, 60 are premature commitments (11.1\%, cluster-bootstrap 95\% CI {[}6.1\%, 16.5\%{]}). C0 wraps the benchmark's stock agent with no gate, and the consistency check in Section 4.2 found no change in mean reward due to the wrapper. We therefore treat the 60 episodes as the behavior of the benchmark agents themselves.

The overall rate conceals a difference between domains. In the airline domain, 58 of 300 C0 episodes (19.3\%) were premature commitments. By model, the rate was 18.0\% {[}8.7, 28.0{]} for gpt-4.1 and 20.7\% {[}12.0, 30.7{]} for gpt-4.1-mini. In the retail domain, only 2 of 240 episodes (0.8\%) were premature commitments, and both models were at 0.8\% {[}0.0, 2.5{]}. The airline policy contains many preconditions under which a request must be refused (basic-economy restrictions, fare-class rules, timing constraints), and the tasks are designed to invite those violations. The retail test split contains comparatively few tasks that require a refusal. This difference in task structure may explain the difference between domains. The reservation-cancellation episode from the introduction is one of the 58 airline C0 cases. Per-cell rates are in \texttt{T4\_boundary\_labels.csv} in the released artifact.

The danger of this phenomenon depends on who evaluates it. Section 5 showed that gpt-4.1-tier judges passed 9 to 11 of the 11 human-confirmed premature commitments, and that the top judge passed 6 of them when given no policy document. The failure type that gates are meant to suppress is thus the one that the judges used in practice detect least well.

\subsection{Do gates change the rate?}

Table 7 compares C0 and C3 in paired form. Every task runs under both conditions, so each task serves as its own control (3 repetitions per condition). Episodes within a task are not independent, so we do not pool episode pairs for the test. Instead, for each task we take the number of successes under C0 minus the number under C3, and the sum of these differences is the test statistic. The null distribution comes from a task-level sign-flip permutation test (the sign of each task's difference is flipped at random; B = 20,000, seed 20260907). This procedure preserves the cluster structure. For each of the 4 cells defined by domain and model, we tested 2 pre-specified metrics: the premature-commitment rate and the overall boundary success rate. Holm correction was applied across the 8 tests \citep{holm}. Discordant-pair counts are shown as descriptive statistics.

\begin{table}[htbp]
\caption{C0 versus C3 by domain and model: discordant-pair counts and task-level sign-flip permutation tests with Holm correction (eight-test family).}\label{tab:perm}
\footnotesize
\setlength{\tabcolsep}{4pt}
\begin{tabular*}{\textwidth}{@{\extracolsep{\fill}}lll>{\raggedright\arraybackslash}p{2.2cm}ll@{}}
\toprule
Domain & Model & Metric & \makecell[l]{Discordant pairs:\\event in C0 only\\/ in C3 only} & \makecell[l]{Permutation $p$} & \makecell[l]{Holm-\\adjusted} \\
\midrule
airline & gpt-4.1 & premature commitment & 9 / 9 & 1.000 & 1.000 \\
airline & gpt-4.1 & boundary success & 21 / 14 & 0.353 & 1.000 \\
airline & gpt-4.1-mini & premature commitment & 11 / 4 & 0.121 & 0.846 \\
airline & gpt-4.1-mini & boundary success & 21 / 20 & 1.000 & 1.000 \\
retail & gpt-4.1 & premature commitment & 1 / 0 & 1.000 & 1.000 \\
retail & gpt-4.1 & boundary success & 17 / 4 & 0.030 & 0.240 \\
retail & gpt-4.1-mini & premature commitment & 0 / 2 & 0.497 & 1.000 \\
retail & gpt-4.1-mini & boundary success & 19 / 10 & 0.150 & 0.902 \\
\bottomrule
\end{tabular*}
\end{table}

After Holm correction, no gate effect was statistically significant. Neither of the two patterns described next is significant. The only cell in which gates tended to help is airline/gpt-4.1-mini. Among discordant pairs, C3 reduced premature commitments from 11 to 4 (permutation p = 0.121). The strongest tendency pointed the other way. In retail/gpt-4.1, C3 turned 17 boundary outcomes that had succeeded under C0 into failures, against 4 turned into successes (permutation p = 0.030, Holm-adjusted p = 0.240). In the domain where Section 6.1 found little to prevent, adding gates tended to make the outcomes worse. Section 6.3 examines, in two analyzed episodes, a gate-implementation defect that can produce this pattern.

We add one behavioral observation. Safe termination under C3 is the path that ends an episode when no policy-compliant action remains. Of the 15 safe terminations observed, 14 came from gpt-4.1 and 1 from gpt-4.1-mini. This suggests the hypothesis that the ability to act and the ability to refrain may grow together, but we did not test it.

\subsection{Trace anatomy: why the retail gate misfired}

Aggregate rates alone do not explain the reversal observed in retail, but the traces do. Figure 3 reconstructs the event sequence of three episodes. Both episodes in Figures 3a and 3b come from the retail/gpt-4.1 cell of Table 7, the cell with the strongest reversal.

\textbf{Figure 3a (retail, C3, gpt-4.1, TB-retail-60-C3-R1@gpt-4.1: false rejection blocks the commit).} The agent drafts a legitimate order-modification write. The gate verifier rejects the draft five times. From the second rejection on, it repeats the same ground, that the order's pending status ``is not confirmed by the conversation.'' The fact was present in the full tool result. When the retry limit is exceeded, the C3 safe-termination path blocks the write and ends the episode. This is a missed action caused by the gate. Native scoring fails the episode; the judge, reading a polite conversation, passes it.

\textbf{Figure 3b (retail, C3, gpt-4.1, TB-retail-32-C3-R2@gpt-4.1: false rejection, then the write is abandoned).} This is a second case of the same defect. The verifier rejects legitimate order cancellation and return requests three times, and the second and third grounds were that ``the customer's identity is not authenticated.'' The authentication step, early in the conversation, was left out of the digest the verifier received. The serialized 20-message digest, which held five long tool results, exceeded 6,000 characters, so the window shrank to 8 messages. After repeated rejections, the agent switches to a response without a write and gives up. No block is recorded; the episode ends as completed, but the required commit is missing. Native scoring fails the episode and the judge passes it. The two episodes show the same root cause surfacing through two different control paths.

\begin{figure}[htbp]\centering
\includegraphics[width=\textwidth]{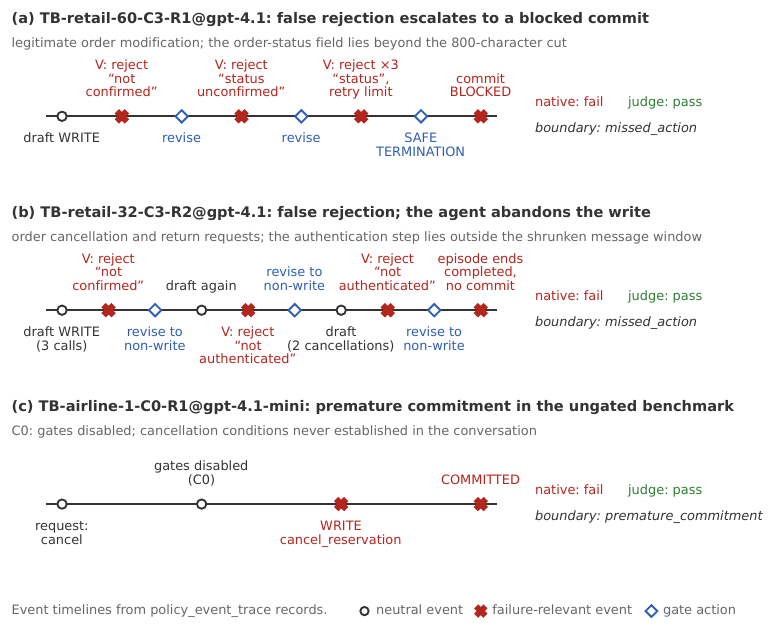}
\caption{Trace anatomy of three episodes, reconstructed from \texttt{policy\_event\_trace} records.}
\end{figure}

\textbf{Figure 3c (airline, C0, gpt-4.1-mini, TB-airline-1-C0-R1@gpt-4.1-mini: premature commitment).} For contrast, this is the phenomenon the gates are meant to block, in its typical form. It is the reservation-cancellation episode from the introduction. A write was executed without the policy's preconditions being met. The judge did not catch it (Section 5); the boundary evaluator did.

The cause of Figures 3a and 3b is a defect in how the gate verifier constructs its evidence. The verifier receives a digest, not the full conversation. The digest covers the most recent 20 messages, keeps only the first 800 characters of each, and shrinks the window to 8 messages when the digest exceeds 6,000 characters (the constants are in Appendix A). Across all 2,160 episodes, airline tool results averaged 663 characters, and 27.5\% exceeded 800. Retail tool results averaged 976 characters, and 59.4\% exceeded 800 (\texttt{tool\_result\_lengths.csv}). Truncation was therefore far more frequent in retail. In the analyzed episodes, each mechanism appeared once. In Figure 3a, the order-status field, at character 950 of a 1,304-character tool result, was cut by the 800-character limit. In Figure 3b, the shrinking of the message window dropped the customer-authentication messages. The verifier then rejected the drafts. Under C1, a failed verification is only recorded and the write proceeds, so the defect does not reach the outcome. Under C2, the agent retries, and if the draft still fails the last draft proceeds, so the commit is not blocked. Only under C3 do repeated verification failures lead to a blocked commit. This mechanism can produce the pattern in Table 7, which appears in the C3 condition and in the retail domain. The mechanism was confirmed in the analyzed episodes; we did not measure how much of the aggregate difference the defect accounts for.

We report the results as pre-specified, without fixing the verifier and re-running. A gate implementation is also software whose quality must be checked separately, and trace-based boundary evaluation made the defect visible and located it. The same class of truncation defect was later found in the annotation-packet builder and was fixed before labeling began (Appendix B).

An independent study of the airline domain of the same benchmark reported that 8 to 17\% of state-changing calls were latent policy violations \citep{nearmiss}. Its definition differs from ours. It counts a violation even when the final state matches the gold state, whereas our premature commitment covers only writes that the gold annotation required the agent to withhold. It nevertheless confirmed the same kind of problem independently on the same benchmark, which is consistent with Section 6.1.

\subsection{Summary}

The results of this section complement the conclusion of Section 5. Section 5 showed that the evaluators measure different things. This section showed that the effect of interventions on agent behavior is not clearly visible, or is misleading, at the aggregate level. The phenomenon itself was concentrated in one domain (Section 6.1). The effect of the gates was not significant after correction for multiple comparisons (Section 6.2). The cause of the reversal was found not in success rates but by following the event sequence (Section 6.3). Whether one is auditing an evaluator or debugging a gate, the unit of evidence that carries explanatory power is the boundary event in the trace.

\section{Implications}

\subsection{The choice of evaluator decides the conclusion}

A counterfactual reading of Table 5 in Section 5 shows the practical meaning of the results. Consider an organization that evaluates agent workflows with an LLM judge, the common choice when no gold annotation exists. It would have concluded that most episodes completed successfully. In our experiment, the gpt-4.1-mini judge passed 92.3\% of the episodes that native scoring failed. It also passed all 11 premature commitments confirmed by human labels. An organization that applied native scoring would have caught the premature commitments. It would have seen only a fail, however, and not the fact that the error occurred at the commit boundary. An organization that applied the boundary evaluator would have seen the error type by name. This holds only where the write expectation was specified in advance. On the policy axis, the evaluator would not have scored above the ``always compliant'' baseline. In the single cancellation case from the introduction, the three verdicts map directly onto three different organizational conclusions. The three evaluators are not imprecise instruments measuring one property. They are separate instruments measuring different properties. Quality assurance for agentic software is therefore not a matter of choosing the most accurate evaluator. It starts with defining the quality property to be measured. Only after clarifying which property a metric measures, and checking that it does, should its numbers be acted on.

\subsection{Three guidelines for practice}

We propose three guidelines for organizations that deploy tool-using agents.

\textbf{First, record commit boundaries at all times.} Commit events are cheap to log. Defining in advance whether each request type expects a write makes the classification deterministic. In our experiment, the errors that gpt-4.1-tier judges mostly missed occurred at the commit boundary. Even the top-tier judge missed more than half of them without the policy document.

\textbf{Second, do not decide a release on an unvalidated judge score alone; run the boundary evaluator alongside it.} The main characteristic of the judges at the gpt-4.1 tier was a consistent false-pass tendency. It persisted when the policy document was provided, when the untruncated input was used, and when a larger model of the same tier was substituted. The top-tier reasoning judge improved only without the policy. Even then it was not significantly better than the boundary evaluator, and with the policy it over-refused (Section 5.3). A judge's strictness is a variable that depends on the model tier and the prompt configuration. It is not a calibrated measurement. The boundary evaluator is deterministic, makes no model calls, and covers the failure type that judges missed most often. It can therefore be added to an existing judge pipeline at low cost. We did not define or evaluate a decision rule that combines the two.

\textbf{Third, budget for the oracle that each evaluator requires.} The cost gap between the boundary evaluator and native scoring is the gap between annotating ``does this request type require a commit?'' and specifying the full goal state. The former is a plausible reliability gate in continuous integration and delivery (CI/CD) that operates on stored traces. The latter has limits as a general practice. One operating structure follows directly from our method. Run and record new episodes at each build, recompute the boundary labels from the stored traces, and route only label changes to human review. This structure requires no model calls at evaluation time.

\subsection{Policy compliance remains open}

The policy axis is the region of Table 5 where results varied least. The weighted agreement of every evaluator fell between 0.595 and 0.672, and none exceeded the ``always compliant'' baseline of 0.688. The top-tier judge dropped to 0.460 when the policy document was included, and raw agreement between the annotators stayed between 0.65 and 0.74. This axis needs further research. A policy-compliance judgment combines document interpretation, counterfactual reasoning about what the agent should have done, and outcome assessment. The human labels converged only after we narrowed the criterion explicitly to outcomes (Section 4.6). Improving this axis requires a structural decomposition of policy compliance, in the same way this paper decomposed the reliability metric. Each policy clause should be converted into a predicate that can be checked against individual events in the trace. The regions open to interpretation should be fenced off separately, with their cost made explicit. Building and validating such clause-level oracles is the main open problem that this study leaves. Clause-level oracles of the kind built by runtime-enforcement systems (Section 2.5) are a candidate, and the labels and traces released with this paper allow them to be tested on the same episodes.

\subsection{Gates must be measured separately}

The Section 6 result, that no gate effect remained significant after correction for multiple comparisons, also carries an implication. The strongest numerical trend ran against the gates. The trace anatomy showed a flaw in how the verifier composed its evidence in the episodes we analyzed. The flaw was an implementation detail, the composition of the evidence digest (Section 6.3). The general point is this. Gate implementations and evaluator implementations are both software whose quality needs separate validation. The same trace format used to evaluate the agent also let us evaluate the gate. It tied the cause of the flaw to a domain-specific interaction, the long outputs of the retail tools. Aggregate rate metrics alone could not have located where this flaw arose.

\section{Threats to validity}

This section sets the scope and the limits of our conclusions. We group the threats into four areas: construct validity, internal validity, external validity, and data integrity and reproducibility.

\subsection{Construct validity}

The human reference labels themselves carry a threat. The second annotator belongs to the same organization as the author. This gives less independence than an annotator from an outside organization would. The labeling procedure is described in Section 4.6. The disagreement observed in the first labeling round (raw agreement 0.650 on axis A2) is consistent with independent judgment. The complete pre-adjudication labels are included in the public materials, so other researchers can recompute the agreement and inspect the adjudication record.

The moderate agreement on axis A2 limits the scope of conclusions drawn from it. We therefore do not state a definite ranking of the evaluators on A2. The sensitivity analysis in Section 5.2 showed that excluding every adjudicated label leaves the relative ranking of the four main evaluators unchanged on A1, A3, and overall. On A2 the ranking changed.

The three-axis label scheme is a simplification of reality. The three axes were judged independently of one another. Rules for exceptional cases were fixed in a version-controlled labeling guideline. These cases include a customer ending the conversation, a failed tool write, and multiple writes in one episode. The remaining room for interpretation is documented in the adjudication record in the public materials.

The boundary evaluator depends on the correctness of its oracle information. Its labels are valid only within the accuracy of the benchmark's annotation of whether a write is expected. The \texttt{correct\_action} label confirms only that a write occurred, not that its content was correct (Section 3.3). Conclusions about content correctness rest only on the native scores and on axis A1.

Implementation flaws in the gate verifier are treated as part of the system under measurement. The evidence-digest flaw observed in the retail domain (Section 6.3) means that the C1 to C3 results reflect the gates as implemented, not ideal gates. We treat this as a subject of empirical analysis in its own right. Its effect on the Section 5 analysis is limited but cannot be ruled out entirely. The comparison among evaluators runs on the same episodes, so it is independent of how the episodes were produced. The distribution of the C1 to C3 episodes in that comparison does, however, reflect an implementation that carried the flaw.

\subsection{Internal validity}

Episodes derived from the same task are not independent (24 per task), and the 120-episode sample was stratified, not a simple random sample. Every inferential procedure therefore accounts for the task-cluster structure. Agreement estimates use a task-cluster bootstrap within domain (B = 10,000) with sampling-weight correction. The C0 versus C3 comparison uses a task-level sign-flip permutation test with Holm correction (Section 6.2).

The C0 consistency check in Section 4.2 showed no systematic performance change from applying the wrapper. It does not show that the wrapper has no effect on behavior.

We state four limits of the judge procedure. First, in the majority vote over two calls, a 1-to-1 tie was treated as a pass. The sensitivity check in Section 4.5 bounded the effect of this rule on the metrics at 0.018. Second, the judge's stochastic variation is bounded by its measured self-agreement (98.2\%), which is small compared with the disagreement structure under analysis. Third, the judge input in our experiment was length-limited. This was examined as a candidate cause for the absence of change in the policy-aware judge. Re-judging with the untruncated input (Section 5.2, Appendix A) produced no significant improvement in weighted agreement. The premature commitments passed by the judge stayed at 10 to 11 of 11. Fourth, the judge results depend on the model tier (Section 5.3). The reasoning model does not support the temperature parameter, so it ran without it, and 11 of its 240 verdicts were ties (Appendix A). The cost totals carry a known error. The per-episode ledger omitted the verifier calls made inside the gates, so the total cost is underestimated by about 13\%. This does not affect the scientific conclusions.

\subsection{External validity}

The benchmark results come from two customer-service domains, one commit of one repository, and two models from the same family and generation. The simulated customer and the gate verifier also used LLMs from the same family. Dual-control domains in which the user executes the action, such as telecom troubleshooting procedures, are outside our scope. The specific figures are not expected to transfer directly to settings outside this one. These include the disagreement rate, the premature-commitment rate, and the cell values of Table 5. What we expect to transfer is the logical structure. This consists of the analysis framework that separates the measured properties, the trace format, and the empirical fact that a widely used evaluator can miss a serious failure type.

The judge results rest on three models of two tiers (two at the gpt-4.1 tier, one top-tier reasoning model) from one vendor and a fixed prompt skeleton. The disagreement analysis over all 2,160 episodes (Section 5.1) used the gpt-4.1-mini judge, and the comparison across model tiers rests on the 120 labeled episodes. Models from other vendors or different prompt configurations may behave differently. The top-tier judge, whose results changed with the information placed in the prompt, is one such case. The traces in the public materials allow the same re-evaluation with other judge configurations.

\subsection{Data integrity and reproducibility}

The risk of data corruption was managed by mechanical checks. Schema validation covered all 2,160 traces and raised no integrity warnings. The analysis pipeline partitions episodes by the pair (storage path, episode ID) to prevent ID collisions between stages. The public materials use collision-free derived identifiers verified by script. Every result table in this paper can be reproduced from the public data and code. The integrity of the labeling rests on three measures. These are blinding, a guideline fixed in advance with a change history, and the preservation of the original labels alongside the final adjudicated labels.

\section{Conclusion}

This paper measured the evaluators with which agentic software is scored. We applied four automatic evaluators to the same 2,160 benchmark episodes and compared them with three-axis human reference labels. The four evaluators were not interchangeable estimates of a single ``success'' metric. They were separate instruments that differed in the information they could access and in the stable direction of their errors. In the single cancellation case from the introduction, native scoring returned a fail, the judge returned a pass, and the boundary evaluator returned premature commitment. This gap between the metrics recurred throughout the 2,160 episodes.

Our main results are the following four.

\textbf{First, judge self-agreement differs from agreement with native scoring.} When the judge re-judged the same episode twice, its self-agreement was 98.2\%. Its agreement with native scoring was only 56.2\%.

\textbf{Second, judge errors are biased.} In the configurations we tested, the direction of a judge's errors changed with the model tier and with the information placed in the prompt. At the gpt-4.1 tier, a false-pass bias dominated regardless of the prompt configuration. The judges passed 9 to 11 of the 11 premature commitments confirmed by human labels. When the policy document was added to the top-tier model, an over-refusal tendency appeared instead.

\textbf{Third, the boundary evaluator did not fall behind the top-tier judge.} The boundary evaluator, which makes no model calls, identified all 11 premature commitments. No judge configuration we tested scored significantly higher than the boundary evaluator on the task-correctness and commitment-correctness axes. The boundary evaluator uses less gold information than native scoring, yet it names the failure type. That diagnosis was used in Section 6 to locate a flaw in the gate implementation.

\textbf{Fourth, premature commitments occur, and the gates have limits.} Premature commitments occurred in 11.1\% of the episodes with the gates disabled. The rate varied more than 20-fold across domains. Enabling the gates produced no significant reduction. Trace analysis identified the loss of evidence in the digest passed to the gate verifier, by a character cut in one episode and by a message-window shrink in another, as a candidate cause of this failure. On the policy axis, no automatic evaluator exceeded the ``always compliant'' baseline.

For software quality assurance practice, the conclusion is not a ranking of individual evaluators but the establishment of an evaluation procedure. Define in advance the quality property that an evaluator measures, validate that fit against a standard independent of the evaluator, and record commit-boundary data at all times. For research, the conclusion is the remaining work. None of the automatic evaluators of policy compliance tested here exceeded a simple baseline, and building one remains an open problem. Follow-up work includes clause-level policy oracles, replication on other benchmark suites and model generations, and extension to dual-control domains. It also includes evaluating learned gate policies in the same trace format used in this paper. The complete traces, source code, full prompts, human labels, and adjudication record used in this study are all released as public materials.

\section*{Declarations}

\bmhead{Funding}

No funding was received for this work.

\bmhead{Competing Interests}

The study was carried out as part of the author's graduate work at aSSIST. The author was employed by QueryPie AI during the study, and the second annotator was a colleague there at the time. No other competing interests are declared.

\bmhead{Data Availability}

The accompanying artifact comprises the gate-injection harness and evaluator implementations, all 2,160 schema-validated traces with derived collision-free identifiers, all evaluator outputs (native, judge, judge-P, boundary), judge and gate prompts with model version identifiers, the annotation kit (rubric v1.3, blinded episode packets, both annotators' original labels, recalibration records, adjudication worksheet with written justifications), the final reference-standard labels, the supplementary material S1--S4 (Online Resource 1), and the analysis code that regenerates every table and figure. The replication package (traces, code, prompts, human labels, and adjudication records) is available at Zenodo, DOI 10.5281/zenodo.23683806 (https://doi.org/10.5281/zenodo.23683806). The tau2-bench repository (commit \texttt{363133a}) is used unmodified under its own license.

\bmhead{Author Contributions}

Sungbin Im designed the study, built the harness and evaluators, ran the experiments, served as the first annotator, performed the analysis, and wrote the manuscript.

\bmhead{Ethics approval and consent}

The study involved no human subjects other than the two annotators, who labeled software execution logs; no personal data were collected. The second annotator gave written consent to be named in the Acknowledgments and to the release of their anonymized original label files as part of the artifact.

\bmhead{Use of AI tools}

Generative AI was used during the preparation of this work in the following ways. Claude (Anthropic) was used for drafting and revising the manuscript text under the author's direction; writing and refactoring the analysis, verification, and packaging scripts released with the artifact; drafting the initial annotation rubric (v1.2); and compiling and checking the reference list. OpenAI Codex was used to review manuscript drafts and to independently re-check the reported numbers against the analysis outputs. Neither tool designed the study, ran the experiments, or produced any human label, adjudication decision, or judgment reported here: all 120 episodes were labeled independently by two human annotators, the recalibration and adjudication were conducted by those annotators, and every reported number was recomputed from the released analysis outputs by scripts the author executed. The author reviewed and takes full responsibility for all content, claims, and code. Large language models are also \emph{objects} of this study (agents, user simulator, gate verifier, and judges); their roles, model identifiers, and prompts are reported in Section 4.4, Appendix A, and supplementary material S1.

\bmhead{Figures}

Figures were produced with Matplotlib.

\bmhead{Acknowledgments}

The author thanks Sungmin Lee, product manager for AI services at QueryPie AI, for serving as the second annotator: independently labeling all 120 episodes on all three axes in the blind round, independently re-judging the policy-compliance axis (A2) after recalibration, participating in the adjudication session, and consenting to the release of the pre-consensus labels that make the agreement statistics in Section 5.2 and Appendix B verifiable. The author also thanks Prof.~Taeyeon Oh, the author's graduate advisor at aSSIST.

\begin{appendices}

\section{Models, judge implementation, and gate configuration}
\setcounter{table}{0}

This appendix reports four items. They are the model identifiers, the construction of the judge input, the re-evaluation procedures with untruncated input and across model tiers, and the configuration of the gate verifier. The verbatim prompts for the judge, the policy-aware judge (judge-P), the gate verifier, the correction feedback, and the safe-termination message are reproduced in supplementary material S1 and in the public materials.

\subsection{Model versions}

Table A1 lists every model used in this study.

\begin{table}[htbp]
\caption{Model identifiers used in the study.}\label{tab:a1}
\footnotesize
\setlength{\tabcolsep}{4pt}
\begin{tabular*}{\textwidth}{@{\extracolsep{\fill}}>{\raggedright\arraybackslash}p{4.4cm} >{\raggedright\arraybackslash}p{8.2cm}@{}}
\toprule
Role & Model identifier \\
\midrule
Agent (gpt-4.1-mini) & \texttt{gpt-4.1-mini-2025-04-14} \\
Agent (gpt-4.1) & \texttt{gpt-4.1-2025-04-14} \\
Simulated customer & \texttt{gpt-4.1-mini-2025-04-14} \\
Gate verifier / correction (C1--C3) & \texttt{gpt-4.1-mini-2025-04-14} \\
Judge and policy-aware judge (main experiment, 2,160 episodes) & \texttt{gpt-4.1-mini-2025-04-14}, temperature 0.4, 2 calls per episode \\
Additional judges (labeled sample, A.3) & \texttt{gpt-4.1-2025-04-14} (temperature 0.4); \texttt{gpt-5.6-sol} (reasoning model; temperature parameter unsupported and omitted; provider-default reasoning effort); 2 calls per episode each \\
\bottomrule
\end{tabular*}
\end{table}

\subsection{Judge input construction and its limits}

The judge's user message has two parts. The first is the customer context, which contains the simulated customer's reason for contact and its scenario instructions. The second is the conversation transcript, which merges the messages by role with the tool calls and tool results. Two length rules apply when the input is built, and we treat them as limits of the judge procedure. Each tool result is included up to its first 600 characters. When the full transcript exceeds 12,000 characters, only its head and tail are kept. The middle is removed and a \texttt{...{[}trimmed{]}...} marker is inserted at the cut. The task gold annotation (\texttt{evaluation\_criteria}) is excluded in every variant. The verdict is parsed from the first JSON object in the response, and an unparseable response counts as a fail with confidence 0. Two independent calls are made per episode and the majority verdict is used. Whether the two calls agree is the self-agreement statistic reported in Section 5.1.

\textbf{Tie rule and sensitivity check.} The two calls can split 1 to 1. The implementation counts this as a pass (\texttt{sum(successes)\ \textgreater{}=\ 1}). Ties occurred in about 2\% of episodes (for example, 49 of 2,168 entries in the judge-P ledger; the ledger holds the 2,160 episodes plus the 8 schema-validation episodes of Section 4.4). This rule favors a pass verdict, and the main result of this study is a false-pass bias. We therefore re-scored the 120 labeled episodes under two alternative rules, using the stored per-call verdicts. The alternatives are (a) a tie counts as a fail, and (b) only the first call counts. No weighted agreement figure moved by more than 0.018. The number of human-confirmed premature commitments passed stayed at 11/11 as in the main text, 10/11 under rule (a), and 11/11 under rule (b).

\subsection{Re-evaluation with untruncated input and comparison across judge tiers}

We re-judged the 120 labeled episodes with both judge variants, using the full input without the tool-result cap and without the transcript trim. The purpose was to test whether judge-P's result was due to the truncation of its input. The model, temperature, prompts, and two-call majority rule were kept the same. The prompt, which tells the judge that it sees the full transcript, was not reworded. Table A2 reports the result. The script (\texttt{judge\_rejudge\_full.py}), the ledger (\texttt{rejudge\_full\_ledger.jsonl}, 240 entries), the per-episode outputs, and the summary table (\texttt{rejudge\_full\_summary.csv}) are included in the public materials. The untruncated transcripts ranged from 2,542 to 26,195 characters (median 11,797), and 57 of the 120 exceeded the original 12,000-character limit. The total cost was USD 0.64. Ties occurred in 4 of the 240 verdicts.

\begin{table}[htbp]
\caption{Re-evaluation of the gpt-4.1-mini judge and judge-P with untruncated input on the 120 labeled episodes.}\label{tab:a2}
\scriptsize
\setlength{\tabcolsep}{1.5pt}
\begin{tabular*}{\textwidth}{@{\extracolsep{\fill}}>{\raggedright\arraybackslash}p{1.6cm}llllllll@{}}
\toprule
Evaluator & A1 & A2 & A3 & Overall & \makecell[l]{False\\pass\\(A1)} & \makecell[l]{False\\fail\\(A1)} & \makecell[l]{Premature\\commitments\\passed} & \makecell[l]{Human A3\\fails\\caught\\(of 53)} \\
\midrule
Judge (main text, length-limited input) & 0.584 & 0.672 & 0.575 & 0.496 & 0.399 & 0.018 & 11/11 & 8 \\
Judge, untruncated & \makecell[l]{0.604\\{}[.508, .700]} & \makecell[l]{0.680\\{}[.582, .768]} & \makecell[l]{0.581\\{}[.485, .677]} & 0.502 & 0.385 & 0.011 & 10/11 & 6 \\
Judge-P (main text, length-limited input) & 0.622 & 0.671 & 0.600 & 0.521 & 0.378 & 0.000 & 11/11 & 9 \\
Judge-P, untruncated & \makecell[l]{0.615\\{}[.521, .709]} & \makecell[l]{0.671\\{}[.573, .763]} & \makecell[l]{0.598\\{}[.501, .694]} & 0.513 & 0.385 & 0.000 & 11/11 & 7 \\
\bottomrule
\end{tabular*}
\end{table}

The paired differences, untruncated minus length-limited (task-cluster bootstrap, B = 10,000), are as follows. For the judge, the change was +0.020 {[}\ensuremath{-}0.017, +0.063{]} on A1, +0.007 {[}\ensuremath{-}0.033, +0.048{]} on A2, and +0.006 {[}\ensuremath{-}0.032, +0.048{]} on A3. For judge-P, the change was \ensuremath{-}0.008 {[}\ensuremath{-}0.035, +0.023{]} on A1, +0.000 {[}\ensuremath{-}0.031, +0.028{]} on A2, and \ensuremath{-}0.001 {[}\ensuremath{-}0.028, +0.029{]} on A3. The verdict changed in 8 episodes for the judge (5 fail to pass, 3 pass to fail) and in 5 episodes for judge-P (4 fail to pass, 1 pass to fail). The conclusions of Section 5.2 hold under untruncated input.

\textbf{Comparison across judge tiers (Section 5.3).} The same untruncated procedure was applied to two stronger models on the 120 labeled episodes (Table A3). The cost was USD 3.06 for gpt-4.1 and USD 6.96 for gpt-5.6 Sol. Ties occurred in 4 of 240 verdicts for gpt-4.1 and in 11 of 240 for gpt-5.6 Sol, which ran without temperature control. The outputs are stored in the public materials as the \texttt{rejudge\_gpt41*} and \texttt{rejudge\_gpt56sol*} files. The script \texttt{taubench\_gates/judge\_models\_analysis.py} regenerates \texttt{analysis\_out/judge\_models\_result.csv} and \texttt{judge\_models\_paired\_diffs.csv}, the source of every number below, from the public CSV files.

\begin{table}[htbp]
\caption{Comparison across judge tiers with untruncated input on the 120 labeled episodes.}\label{tab:a3}
\scriptsize
\setlength{\tabcolsep}{1.0pt}
\begin{tabular*}{\textwidth}{@{\extracolsep{\fill}}>{\raggedright\arraybackslash}p{1.9cm}llllllll@{}}
\toprule
\makecell[l]{Judge\\(untruncated\\input)} & A1 & A2 & A3 & Overall & \makecell[l]{False\\pass\\(A1)} & \makecell[l]{False\\fail\\(A1)} & \makecell[l]{Premature\\commitments\\passed} & \makecell[l]{Human A3\\fails\\caught\\(of 53)} \\
\midrule
gpt-4.1-mini, no policy & \makecell[l]{0.604\\{}[.508, .700]} & \makecell[l]{0.680\\{}[.582, .768]} & \makecell[l]{0.581\\{}[.485, .677]} & 0.502 & 0.385 & 0.011 & 10/11 & 6 \\
gpt-4.1-mini, policy & \makecell[l]{0.615\\{}[.521, .709]} & \makecell[l]{0.671\\{}[.573, .763]} & \makecell[l]{0.598\\{}[.501, .694]} & 0.513 & 0.385 & 0.000 & 11/11 & 7 \\
gpt-4.1, no policy & \makecell[l]{0.617\\{}[.512, .719]} & \makecell[l]{0.667\\{}[.567, .759]} & \makecell[l]{0.603\\{}[.499, .705]} & 0.543 & 0.331 & 0.053 & 9/11 & 14 \\
gpt-4.1, policy & \makecell[l]{0.615\\{}[.513, .715]} & \makecell[l]{0.666\\{}[.570, .758]} & \makecell[l]{0.593\\{}[.490, .694]} & 0.542 & 0.340 & 0.044 & 10/11 & 10 \\
gpt-5.6 Sol, no policy & \makecell[l]{\textbf{0.737}\\{}[.650, .819]} & \makecell[l]{0.599\\{}[.506, .694]} & \makecell[l]{\textbf{0.673}\\{}[.583, .759]} & 0.685 & 0.131 & 0.132 & 6/11 & 34 \\
gpt-5.6 Sol, policy & \makecell[l]{0.580\\{}[.482, .671]} & \makecell[l]{0.460\\{}[.360, .563]} & \makecell[l]{0.557\\{}[.466, .647]} & 0.647 & 0.000 & 0.420 & 0/11 & 51 \\
\emph{Boundary evaluator (reference)} & 0.766 & 0.595 & 0.708 & 0.664 & 0.216 & 0.019 & 0/11 & 24 \\
\emph{Native scoring (reference)} & 0.818 & 0.653 & 0.758 & 0.766 & 0.070 & 0.112 & 0/11 & 40 \\
\bottomrule
\end{tabular*}
\end{table}

The paired differences (task-cluster bootstrap) are as follows. The differences between gpt-4.1 and gpt-4.1-mini were within \ensuremath{\pm}0.022 on every axis, and every confidence interval included zero. gpt-5.6 Sol (no policy) versus gpt-4.1-mini: A1 +0.133 {[}+0.016, +0.250{]}, A3 +0.091 {[}\ensuremath{-}0.024, +0.210{]}. gpt-5.6 Sol (no policy) versus the boundary evaluator: A1 \ensuremath{-}0.029 {[}\ensuremath{-}0.140, +0.082{]}, A3 \ensuremath{-}0.036 {[}\ensuremath{-}0.146, +0.078{]}, overall +0.021 {[}\ensuremath{-}0.086, +0.134{]}. gpt-5.6 Sol (no policy) versus native scoring: A1 \ensuremath{-}0.081 {[}\ensuremath{-}0.164, \ensuremath{-}0.002{]}, A3 \ensuremath{-}0.086 {[}\ensuremath{-}0.174, \ensuremath{-}0.002{]}. gpt-5.6 Sol with the policy versus without it: A1 \ensuremath{-}0.156 {[}\ensuremath{-}0.270, \ensuremath{-}0.041{]}, A3 \ensuremath{-}0.116 {[}\ensuremath{-}0.235, +0.004{]}. Of the 55 episodes that the annotators judged correct on all three axes, gpt-5.6 Sol failed 11 without the policy and 40 with it. gpt-4.1-mini with the policy failed none of them.

Against the two reference evaluators, every untruncated judge configuration was significantly below native scoring on A1 and A3 (gpt-4.1-mini: A1 \ensuremath{-}0.214 / \ensuremath{-}0.204, A3 \ensuremath{-}0.177 / \ensuremath{-}0.160; gpt-4.1: A1 \ensuremath{-}0.202 / \ensuremath{-}0.203, A3 \ensuremath{-}0.155 / \ensuremath{-}0.166; gpt-5.6 Sol with policy: A1 \ensuremath{-}0.238, A3 \ensuremath{-}0.202; no-policy / policy order; all confidence intervals exclude zero). On A2, no configuration differed significantly from native scoring except gpt-5.6 Sol with the policy (\ensuremath{-}0.194). Against the boundary evaluator, the untruncated gpt-4.1-mini configurations were significantly lower on A1 (\ensuremath{-}0.162 / \ensuremath{-}0.151) and A3 (\ensuremath{-}0.127 / \ensuremath{-}0.110). gpt-4.1 was significantly lower on A1 (\ensuremath{-}0.149 / \ensuremath{-}0.150) and, with the policy, on A3 (\ensuremath{-}0.116; without the policy \ensuremath{-}0.105 {[}\ensuremath{-}0.212, +0.004{]}). gpt-5.6 Sol with the policy was significantly lower on A1 (\ensuremath{-}0.185) and A3 (\ensuremath{-}0.152). The full table is \texttt{analysis\_out/judge\_models\_paired\_diffs.csv}.

\subsection{Gate configuration and verifier evidence digest}

The gate verifier ran on \texttt{gpt-4.1-mini-2025-04-14} at temperature 0.0, with at most 2 regeneration attempts per rejected write. The domain policy document is attached inside \texttt{\textless{}policy\textgreater{}} tags in the system prompt. The verbatim prompt, the correction feedback injected after a rejection, and the C3 safe-termination message are reproduced in supplementary material S1.3.

\textbf{Verifier evidence digest (the mechanism analyzed in Section 6.3).} The verifier does not see the raw conversation. It receives a digest built from the most recent 20 messages. Each message is kept up to its first 800 characters. When the serialized digest exceeds 6,000 characters, the window shrinks to the most recent 8 messages. Tool results in the retail domain average 976 characters, and 59.4\% of them exceed 800 characters. In the airline domain, the average is 663 characters and 27.5\% exceed 800. The 800-character cut of tool results therefore occurred far more often in retail. In the analyzed episodes, the cut and the window shrink each removed a key precondition from the verifier's evidence (Section 6.3). This contributed to the false rejections discussed in Section 6.3. These constants are part of the measured system, so we publish them with it.

\section{Labeling rubric, adjudication procedure, and agreement computation}
\setcounter{table}{0}

This appendix summarizes how the human reference labels were produced. The public materials contain the complete version-controlled guideline, both annotators' original label files, the recalibration records, and written justifications for every adjudication decision.

\subsection{Rubric structure (v1.2 and the v1.3 amendment)}

The labeling guideline applies the following rules in order.

\begin{itemize}

\item
  Evidence principle: a judgment rests only on the domain policy document and the observed evidence provided. It does not rest on prior expectations about how agents usually behave.
\item
  Axis independence: the three axes are judged separately. The guideline includes a table of episodes in which the axes diverge (for example, a request confirmed correctly but applied to the wrong item is A1 ``no'', A2 ``yes'', A3 ``no'').
\item
  Per-axis criteria: the guideline defines the ``yes'', ``no'', and ``not applicable'' criteria for each axis and the handling of boundary cases. Examples are an episode that ends with a transfer to a human agent, and a customer who withdraws the request.
\item
  Three-step judgment order: check what the policy permits, then check whether the preconditions are met, then compare the executed actions against the observed facts.
\item
  Exception rules: five rules cover termination markers, a customer ending the conversation before the agent could act, a failed tool write, an ending by transfer to a human agent, and multiple writes in one episode.
\end{itemize}

Two amendments were made under version control, and both are preserved in the public materials.

\begin{itemize}

\item
  v1.3 amendment (criteria-alignment meeting, before the second A2 round): the A2 policy-compliance axis is judged on the state at the end of the conversation. An episode that corrected its own mid-conversation error counts as compliant.
\item
  Supplement to the customer-termination rule (added at the adjudication session): a state-changing request that the customer has fully confirmed remains a task the agent must carry out. This holds even if the customer then ends the conversation. A confirmed but unexecuted request is therefore A3 ``no''. This supplement was applied uniformly to the 7 episodes that fall under it, each with a written justification.
\end{itemize}

\subsection{Procedure}

\begin{enumerate}
\def\labelenumi{\arabic{enumi}.}

\item
  Rubric freeze: the annotators were trained on one out-of-sample practice episode, and the guideline was frozen.
\item
  Round 1, independent: both annotators labeled all 120 episodes on the three axes independently and blind.
\item
  Agreement computation: agreement on axis A2 came out low (raw agreement 0.650).
\item
  Pre-declared recalibration: following the procedure fixed in advance, one joint criteria meeting produced guideline v1.3. Both annotators then re-labeled axis A2 only, independently, on all 120 episodes. No repeated rounds were run to reach a target value.
\item
  Adjudication: every disagreement remaining after the two rounds (62 episodes, 80 axis decisions) was resolved in an evidence-based adjudication meeting. The original labels were preserved. An unresolved outcome was permitted, but none occurred.
\item
  Reference freeze: the final reference labels were frozen. Every statistical analysis that followed (Section 5.2) ran after this freeze.
\end{enumerate}

\subsection{Agreement statistics}

Table B1 reports agreement by axis and round.

\begin{table}[htbp]
\caption{Inter-annotator agreement per axis and round (two annotators, 120 episodes).}\label{tab:b1}
\small
\setlength{\tabcolsep}{4pt}
\begin{tabular*}{\textwidth}{@{\extracolsep{\fill}}lllll@{}}
\toprule
Axis & Round & Raw agreement & \makecell[l]{Cohen's $\kappa$} & \makecell[l]{Gwet's AC$_1$} \\
\midrule
A1 & 1 & 0.842 & 0.686 & 0.685 \\
A2 & 1 & 0.650 & 0.317 & 0.315 \\
A2 & 2 (after recalibration) & 0.742 & 0.306 & 0.595 \\
A3 & 1 & 0.750 & 0.477 & 0.526 \\
\bottomrule
\end{tabular*}
\end{table}

For two annotators and binary labels over \(n\) episodes, let \(p_a\) be the observed agreement. Cohen's \(\kappa = (p_a - p_e)/(1 - p_e)\), where \(p_e\) is the chance agreement computed from the two annotators' label distributions \citep{cohen}. Gwet's \(\mathrm{AC}_1 = (p_a - p_\gamma)/(1 - p_\gamma)\), where \(p_\gamma = 2\pi(1-\pi)\) and \(\pi\) is the mean of the two annotators' ``yes'' rates \citep{gwet}. \ensuremath{\kappa} is vulnerable to the prevalence paradox. When the annotators' labels concentrate on one value, \(p_e\) grows, and \(\kappa\) comes out low even when the observed agreement \(p_a\) is high. We therefore pre-specified AC1 alongside \ensuremath{\kappa} before the study. The second A2 round (raw agreement 0.742, AC1 0.595, \ensuremath{\kappa} 0.306, both annotators' ``yes'' rates at or above 0.69) shows this property of the two statistics. The computation script is included in the public materials and reproduces these values from the public label files.

\subsection{Blinding and integrity}

The labeling packets contained the domain policy, the customer context, the full transcript, the state-changing tool calls with their results, and the observed commit facts. They excluded the gate condition, the model used, the verdicts of the four automatic evaluators, the task gold annotation, and the mapping between traces and episode IDs. That mapping was kept in a separate file never shown to the annotators. An audit during packet preparation found a tool-result truncation defect (the same class as in Section 6.3). The defect was fixed and all packets were regenerated. The audit record and a list of the effects of each guideline change are included in the public materials. The original pre-adjudication labels are preserved separately from the final adjudicated labels, so every agreement statistic in this paper can be recomputed from the primary data.

\end{appendices}

\bibliography{sn-bibliography}
\end{document}